# On a possibility of transfer of asteroids from the 2:1 mean motion resonance with Jupiter to the Centaur zone

Kazantsev A. M., Kazantseva L. V.

Astronomical Observatory of Kyiv Taras Shevchenko National University, Ukraine
ankaz@knu.ua

**ABSTRACT**
The paper analyses possible transfers of bodies from the main asteroid belt (MBA) to the Centaur region. The orbits of asteroids in the 2:1 mean motion resonance (MMR) with Jupiter are analysed. We selected the asteroids that are in resonant orbits with $e > 0.3$ whose absolute magnitudes $H$ do not exceed $16^m$. The total number of the orbits amounts to 152. Numerical calculations were performed to evaluate the evolution of the orbits over 100,000-year time interval with projects for the future. Six bodies are found to have moved from the 2:1 commensurability zone to the Centaur population. The transfer time of these bodies to the Centaur zone ranges from 4,600 to 70,000 yr. Such transfers occur after orbits leave the resonance and the bodies approach Jupiter. Where after reaching sufficient orbital eccentricities bodies approach a terrestrial planet, their orbits go out of the MMR. Accuracy estimations are carried out to confirm the possible asteroid transfers to the Centaur region.

## 1. INTRODUCTION

Centaurs are small bodies moving between the orbits of Jupiter and Neptune. As some of them exhibit comet properties, Centaurs are treated as an important source of the Jupiter family comet nuclei (Horner et al. 2004). Most scientists regard the Kuiper belt as the origin area of Centaurs with the scattered disk being their main source (Levison & Duncan 1997; Di Sisto & Brunini 2007; Volk & Malhotra 2008). Furthermore, additional sources thereof may be the bodies in the MMR with Neptune, 3:2 (Morbidelli 1997; Di Sisto et al. 2010), and 2:1 (Tiscareno & Malhotra 2009). When found in orbits with high inclinations ($i > 70°$) or in orbits with retrograde motion ($i > 90°$), Centaurs are considered as coming from the Oort cloud or from the interstellar space (Brasser et al. 2012; Namouni & Morais 2020). Neptune and Jupiter Trojans may be deemed as another two additional minor sources of Centaurs (Horner & Lykawka 2010; Di Sisto et al. 2019), whereas inner zones of the Solar system are not typically regarded as possible Centaur sources.

From our perspective, the MBA may also contribute to building up the Centaur numbers. Some MMR with Jupiter are found in the MBA. Their orbit eccentricities may reach certain values that are sufficient for the bodies to approach the planets. Where such an approach occurs, the semimajor axes of asteroid orbits can undergo significant changes, and the bodies can move to other areas of the Solar system, which explains why Kirkwood gaps exist in the vicinity of the MBA resonances. The most notable gaps are observed in resonances: 3:1, 5:2, and 2:1. The ν6 secular resonance, the Mars-crossing population, and the 3:1 MMR with Jupiter are the main sources of near-Earth asteroids (NEAs). The 5:2 and the 2:1 resonances, which are located in the outer belt, are minor NEAs sources (Bottke et al. 2002). When approaching the planets, small bodies can move farther to more remote areas of the Solar system. On average, over a long time, increases in semimajor axes have outweighed decreases in such approaches (Kazantsev, 2012). Thus, the asteroid belt, in particular in the 2:1 resonance zone, may also be a source that replenishes the Centaur numbers.

## 2. THE CALCULATION METHOD

The main results have been obtained on the basis of numerical integration of the orbit evolutions of the Solar system bodies. Therefore, the integration method it worthwhile mentioning here. Its detailed description is found in Kazantsev (2002).

Integrated equations are those for second time derivatives of coordinates. The $x$, $y$ and $z$ coordinates are represented in the rectangular heliocentric system. In contrast to the classical variant, we integrate only second time derivative components caused by disturbances. The total gravitational influence of the Sun is calculated by the formulas of undisturbed motion. In so doing, we deem it necessary to apply the transition from coordinates to orbit elements (and vice versa) at every integration step, which is the main disadvantage of this approach. While half a century ago, before computer technologies became widespread, such a method of calculations was not possible, it poses no problem nowadays.

Simpson's rule is used to integrate the equations proper. In the middle of the integration step and at its end, the required coordinate values are determined by Taylor series up to and including 3-rd derivatives. At the outset, three applications were written for integration: one was meant only Taylor's method for 4-th derivative inclusive; the other one relied on the 5-th order Runge-Kutta method; and the Simpson's rule was used in the third one. When checked for efficiency, each of the programmes showed that they are all very close where accuracy and time are calculated. As the variant with the Simpson's rule is the most handy to describe, we decided to choose it. We did not compare it with the other methods of integration.

In recent years, the Bulirsch-Stoer method has been widely used. An evaluation of this method effectiveness can be found in Murison (1989). This paper shows that the Bulirsch-Stoer integration requires less computer time than the 4-th order Runge-Kutta method. However, integration by the Bulirsch-Stoer method yields greater errors where a small body approaches a planet, than those that arise where the 4-th order Runge-Kutta method is applied. As our method is not less effective than the 5-th order Runge-Kutta method, and as far as major errors occur when asteroids approach the planets, we do not believe that our method of integration should be replaced by another one.

The main advantage of this approach over the classical one is the minimal value of the integrable function. This allows for choosing a much larger integration step, which increases the accuracy of calculations at long-term intervals.

The calculation accuracy has been repeatedly tested in a number of ways. The accuracy over short time intervals (decades) was checked by calculating individual body positions in the past. The comparison of these positions with the data on the MPC website showed that the accuracy of the calculations is not worse than the accuracy of the observations. This applies not only to asteroids, but also to more distant bodies, in particular, Pluto.

Nevertheless, to estimate the calculation accuracy of long-term intervals, one needs another approach. Such calculations can be conventionally termed forward and backward computations. To do so, we perform numerical calculations in a certain time direction with an interval (up to $t_k$). In the same calculation process, without stopping the calculations, we perform the reverse process, i.e. calculate the orbit backward to the initial moment $t_0$. The difference between the initial and calculated elements of the orbit at the time $t_0$ characterises the calculation error. While the direction of the integration changes, the errors in the calculations that have accumulated up to the moment $t_k$ continue to grow. We get errors in the interval $2t_k$ when we return to the moment $t_0$.

While calculating forward and backward, our application found the calculation error for the semimajor axes of the asteroid orbits to be $1 \times 10^{-7}$ AU, whereas for the eccentricities it constitutes $1 \times 10^{-7}$ (the accuracy of the MPC catalogue) at intervals up to 50,000 yr. This accuracy is maintained for different MBAs in the absence of approaches to the planets.

Yet, this technique allows us to check only the internal accuracy of the method. The real (absolute) accuracy of calculations depends on the integration step that we choose and way it changes in the calculation process. This is particularly true for calculating orbits of small bodies that are in close approaches to the planets, first of all giant planets. Approaches to giant planets significantly change orbit elements. In most cases, the integration of such orbits can be considered accurate until the first close approach to the planet takes place. To assess the absolute accuracy of calculations, the following method can be used. After completing the calculation of the orbit evolution with a certain initial step *sp*1, one needs to recalculate the same orbit with a significantly smaller initial step where *sp*2 = *sp*1/2. If the results match, they can be considered accurate.

Perturbations from all eight major planets (Mercury – Neptune) were taken into account in the present calculations. All asteroids were treated as massless particles. The integration step was chosen to be 2 days long. As an asteroid approached a planet, the step size decreased depending on the distance between the asteroid and the planet.

The relativistic effects of orbital perihelion displacement were taken into account as well. To this end, we used the widely known formula,

$$\Delta\omega = 6 \times \pi \times G \times M_s / (c^2 \times a \times (1 - e^2))$$

where $\Delta\omega$ is a relativistic shift of the argument of the body orbit perihelion for one revolution around the Sun (in radians), $\pi$ is a mathematical constant, $G$ – gravitation constant, $M_s$ – Solar mass, $c$ – light velocity, $a$ and $e$ – the semimajor axis and eccentricity of the body orbit.

The reason why such effects should be taken into account will be described below.

## 3. CALCULATIONS AND THEIR RESULTS

### 3.1 Features of the small body motion in commensurabilities with planets

Every commensurability is characterised by a certain value of the semimajor axis $a_c$. If the semimajor axis of the asteroid orbit equals $a_c$, then the mean motion of the planet $n_p$ and the mean motion of the small body $n_a$ correlate as integers:

$$n_p/n_a = p/q \qquad (1)$$

This ratio is called mean motion resonance. Semimajor axes of small body orbits periodically oscillate in regard to the $a_c$ value, in the proximity of the commensurability. Therefore, for all orbits in commensurability, equation (1) holds only approximately.

For each value of $a_c$ there are stable and unstable resonant zones determined by a set of three angular elements, namely: the argument of the perihelion $\omega$, the ascending node longitude $\Omega$ and the mean anomaly $M$. In the 1:1 resonance, for instance, stable zones are located near the Lagrange points $L_4$ and $L_5$. Bodies in stable zones are always at safe distances from planets.

The orbit conditions in stable resonant zones can be determined by a certain angular value $\sigma$, called the libration argument. Proposed by Schubart (1968), this parameter is described by the expression

$$\sigma = q \times (\pi_p + M_p) - p \times (\pi_a + M_a) + (p - q) \times \pi_a, \qquad (2)$$

where $\pi_p$ and $\pi_a$ are longitudes of the planet orbit perihelion and of the asteroid; $M_p$ and $M_a$ are mean anomalies of these bodies; $p$ and $q$ are values from the expression (1). In a stable resonant zone, the libration argument performs periodic oscillations with a certain amplitude *ds*, i.e. it librates. The smaller the amplitude of oscillations of the $\sigma$ value, the greater the distance of the asteroid from the planet, and the more stable is its resonance. Formally, the value of *ds* can be within ± 180°. However, where *ds* values are sufficiently high, the danger for this orbit to fall out of the resonance also increases. In addition to the gravity of the planet, some other contributing factors may exist. These may be gravitational influences of other bodies and collisions with other asteroids. They are able to force the orbit out of the stable resonant zone.

For orbits outside the resonance, the value of σ varies in the entire angular range from 0° to 360° (it circulates). Bodies in such orbits, even when uninfluenced by other factors, can approach a planet creating resonance.

**3.2 Resonant orbit selection in the 2:1 commensurability**

The zone of the 2:1 commensurability with Jupiter is located on the outer edge of the MBA ($a_c$ = 3.278 AU). Fig. 1 shows the distribution of asteroid orbits in the coordinates of the semimajor axis – eccentricity ($a - e$) in the vicinity of this commensurability. Here, asteroids from the MPC catalogue of April 27, 2019 with absolute magnitudes $H < 16^m$ are selected. The figure shows a certain rarefaction in the numbers of bodies in the vicinity of commensurability. Also, it is obvious that in the commensurability the orbit eccentricities can reach quite high values. In this project we also analysed orbits of smaller asteroids ($H < 18^m$). However, in order not to overburden Fig.1 with minor details, they are left out.

Bodies in orbits with large eccentricities can leave commensurability when they approach the planets. Asteroids in orbits with $e > 0.3$ were selected for further analysis. The number of such bodies with $H < 18^m$ exceeded 600.

From this array, resonant orbits were selected for further analysis, as we are mainly concerned with the issue whether bodies just from the 2:1 MMR with Jupiter may transfer to the Centaur region.

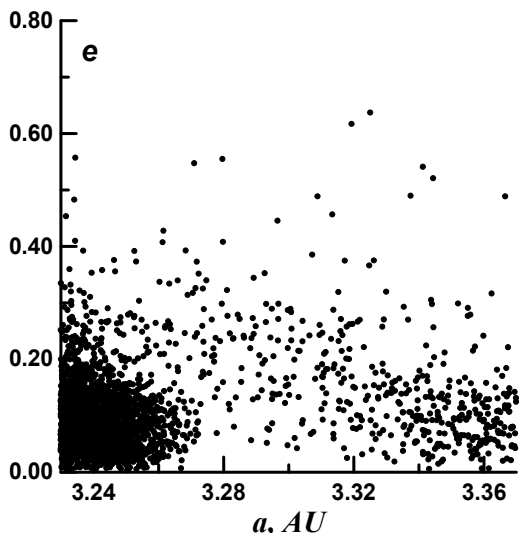 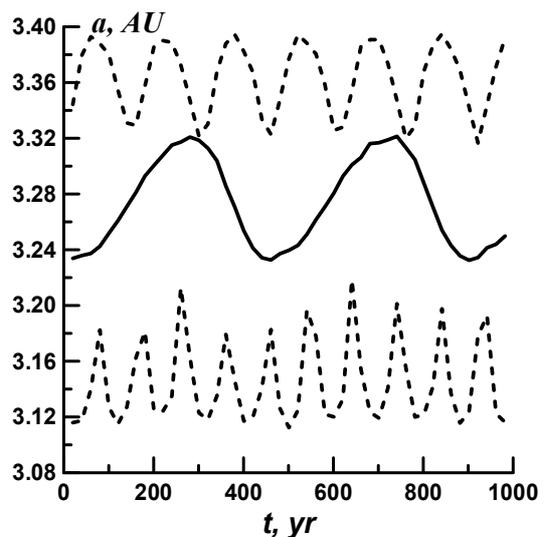

*Figure 1.* The $a - e$ distribution of orbits in the 2:1 commensurability with Jupiter (absolute magnitude $H < 16^m$)

*Figure 2.* The $a(t)$ dependence for an resonant orbit (solid line) and for non-resonant ones (dotted lines)

The detection of resonant orbits was based on numerical calculations of evolution over short time intervals (up to 1,000 yr). At such intervals, it is already obvious whether such an orbit is resonant or not. The semimajor axes of the resonant orbits periodically intersect the values of $a_c$ (3.28 AU) in the course of the evolution, in contrast to the non-resonant orbits (Fig. 2). It was the $a(t)$ dependence that defined the choice of resonant orbits. As the σ value librates for all resonant orbits and circulates for non-resonant ones, the choice can be made using σ(t).

As a result of preliminary numerical calculations, 152 resonant orbits were detected.

## 3.3 Numerical calculations of the resonant orbits

Numerical calculations of these resonant orbit evolutions were performed at intervals up to 100,000 yr. Our calculations showed that where bodies were in the 2:1 commensurability with Jupiter, they could pass into Centaurs. Such transitions occurred after the orbit left the resonance and the body approached Jupiter. The exit from resonance is probably due to approaches to the planets of the terrestrial group. In most cases, where such approaches occurred, the semimajor axes of the orbits remained almost unchanged. But after such an approach, the asteroid orbit may leave the stable resonance zone. After that, the body may approach Jupiter. In total, 6 bodies passed to Centaurs. Those were asteroids: 322713, 406803, 451124, 471179, 475265, and 506437.

The other 146 orbits remained in resonance or nearby. Eleven orbits left the resonance to enter it again. Those were the orbits with either $\sigma > 160°$, or $e < 0.05$. Where the eccentricity was small, the resonance stability decreased.

In the course of the orbit evolutions, 23 bodies out of 146 moved to the NEA population for varying spans of time, with 11 bodies belonging to the NEAs from the outset. We imply here that bodies belong to the NEAs if their perihelion distances $q$ are smaller than 1.3 AU. For 13 orbits the perihelion distance fell down to $q < 1.0$ AU. The issue of the NEA population replenishment from resonance 2:1, however, should be considered in a separate study.

As of today, the ranges of the Centaurs orbit elements have not been clearly defined yet. If we consider Centaurs to be bodies that are constantly found in the space between the orbits of Jupiter and Neptune, then we need to select the orbits which have perihelion distances $q > 5.2$ AU, and aphelion distances $Q < 30$ AU. If Centaurs can enter the space between the orbits of the giant planets, at least for a while, then there should be two groups of orbits: 1) those with $q < 30$ AU, 2) those with $Q > 5.2$ AU. We adopted an intermediate variant, which is often applied as well, where the orbit semimajor axes are $5.2 < a < 30$ AU.

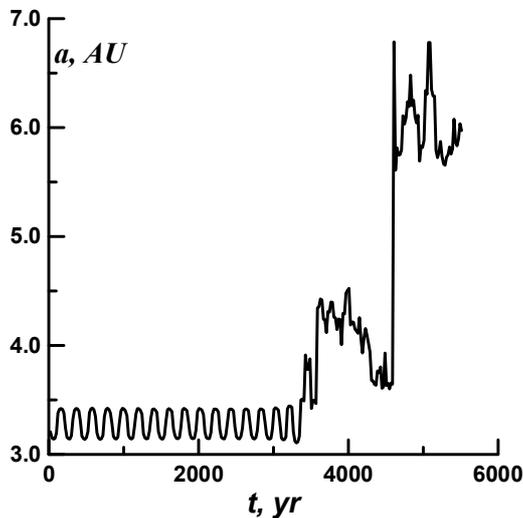

*Figure 3.* The $a(t)$ dependence for the asteroid №406803

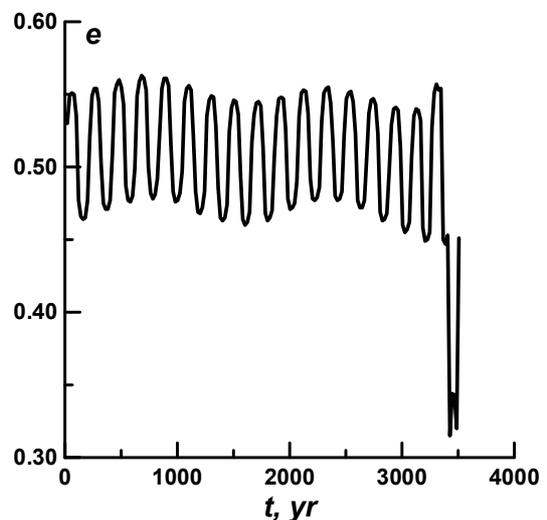

*Figure 4.* The $e(t)$ dependence for the asteroid №406803

The transfer of asteroid №406803 to Centaurs is worth examining in more detail. The $a(t)$, $e(t)$ and $\sigma(t)$ dependences for this asteroid are presented in Figs 3-5. The latter two dependences are shown before the orbit exits the resonance.

The orbit was in resonance for less than 3,500 years. The asteroid frequently approached Mars. The $ds$ value changed little during that time. But after the asteroid approached Mars again, ($t = 3280$ yr), the $ds$ value increased from 152° to 170°, and then the orbit went out of the resonance. In 130-year time, there was an approach to Jupiter, and the body went far beyond the 2:1 commensurability. After that, the asteroid approached Jupiter only. After reaching the distance of 0.02 AU (*t* = 4,600 yr) the asteroid entered the Centaur zone.

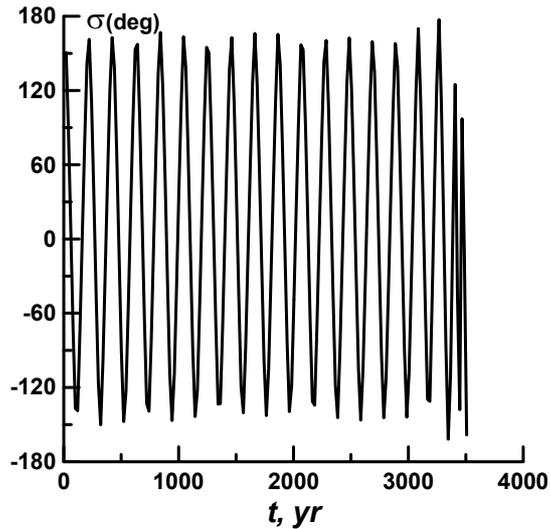

*Figure 5.* The σ(*t*) dependence for the asteroid №406803

The calculation accuracy is definitely crucial for these results. It depends on two key factors: the accuracy of the method per se, and the accuracy of the initial orbit elements, namely: of the major planets and small bodies. In the MPC catalogue, for numbered asteroids, the accuracy of orbit elements is as follows: the semimajor axes – $10^{-7}$ AU, the eccentricities – $10^{-7}$, and the angular elements – $10^{-5}$ degree. Such accuracy for orbit elements may be achieved after the asteroid has been observed in several oppositions. All numbered asteroids, including those that passed to Centaurs, meet these prerequisites. In particular, for more than 20 years, asteroid №406803 was observed in 5 oppositions. The total number of the positions obtained is 251.

In calculations, the total error accumulates at every integration step. The smaller is the step, the higher is the accuracy of the results at a single step. If the number of steps increases, the total error builds up. Therefore, calculations for long time intervals can hardly be considered accurate, especially if there are approaches to the planets.

The clone method is quite often used to estimate the evolution of an orbit over long time intervals. The essence of the method consists in creating a number of artificial models (clones) of the one orbit. Clone orbits differ insignificantly from real ones in individual orbit elements, the main reason being that the process of a real orbit development occupies an intermediate position among that of clone orbits. This approach is mainly used to estimate the life time of a body in a population (Horner et al. 2004; Di Sisto et al. 2010). But it is definitely impossible to determine the correct course of the evolution of a real orbit with the help of clones.

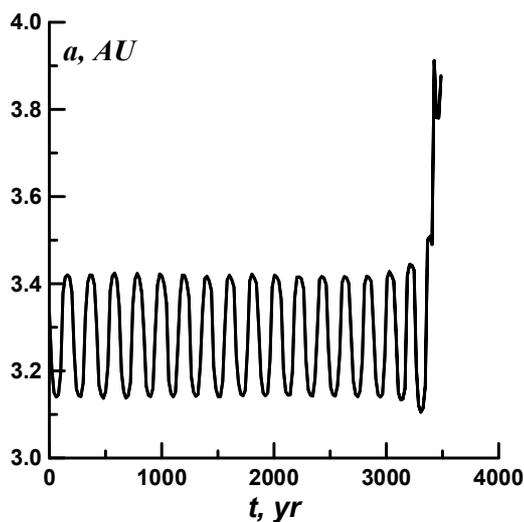

*Figure 6.* The *a*(*t*) dependence for the asteroid №406803, calculated forward and backward at the interval of 3500 yr

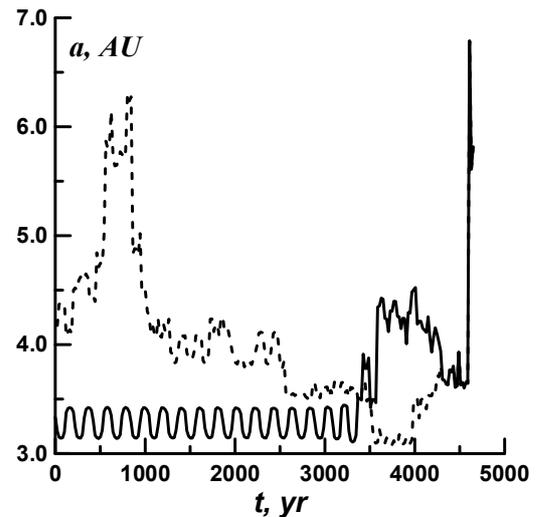

*Figure 7.* The *a*(*t*) dependence for the asteroid №406803, calculated forward and backward at the interval of 4600 yr

To evaluate accuracy of numerical integration results of an orbit, we used forward and backward computations, and integrations with two different steps. In particular, such estimates were made for all cases of transition to the Centaur region.

Fig. 6 shows the orbit evolution of asteroid №406803 calculated forward and backward at the interval of 3,500 yr. In the Figure, the courses of the forward and backward evolution merge into one line. The difference between the initial and calculated values of the $a$ at the time $t_0$ does not exceed 0.001 AU. The calculations with steps of 2 days and 1 day also showed a difference smaller than 0.001 AU. Therefore, the orbit evolution calculations to 3,500 yr can be considered accurate. It means that after 3,500 yr the orbit went out of resonance. The semimajor axis reached the value of 3.90 AU, with the eccentricity being 0.32. The orbit did not fall in the 3:2 MMR with Jupiter.

If we perform forward and backward calculations up to the transition to the Centaur region (4,600 yr), the forward and backward evolution courses will significantly differ (Fig. 7). The backward course is marked by the dotted line. The Figure demonstrates a limited calculation accuracy where asteroids repeatedly approach Jupiter. However, the transition of this asteroid to the Centaur region is unlikely to be doubtful. After all, the accuracy of forward and backward calculations is confidently confirmed up to the point where the asteroid comes out of resonance and until it first approaches Jupiter. After that, the transition of the asteroid to the Centaur region is only a matter of time.

Out of the six bodies that reached Centaurs, apart from asteroid №406803, for one more of them (№322713), the calculations dealing with the body coming out of the resonance and first approaching Jupiter can be considered quite accurate. The other four asteroids that passed to Centaurs stayed in the 2:1 commensurability much longer. During all that time, the bodies had experienced numerous approaches to the planets of the terrestrial group. Due to that motion, though the bodies did go out of the resonance, forward and backward calculations cannot be deemed sufficiently accurate yet.

The differences between the initial values of $a$ at $t_0$ moment and the values calculated forward and backward, until the moment of escaping from resonance for those four orbits, constitutes from 0.2 AU (№506437) to 5 AU (№471179). Orbit №471179 was in resonance for the longest period of time (63,000 yr); therefore, the difference here was the largest. The course of its evolution computed forward and backward until the moment it exits the resonance resembles the course of the orbit №406803 evolution before its transition to the Centaur region (Fig. 7). For orbit №406803, errors in the elements began to increase significantly after first approaching Jupiter. For those four orbits, errors had gradually accumulated over a long evolution, as the bodies approached the terrestrial planets.

Nevertheless, the asteroid orbit evolution calculations in the 2:1 commensurability with Jupiter indicate quite clearly that some part of these bodies are very likely to have moved to the Centaur region. The main conditions for such movements are the significant eccentricities achieved by the orbits and their departure from the stable resonance zone, which are the prerequisites for the bodies to approach Jupiter. It is difficult to estimate quantitatively the transition of bodies from the 2:1 commensurability to the Centaur region. Ultimately, the MPC 2019 catalogue is very unlikely to hold a complete sample of bodies with the absolute magnitude $H < 18^m$ in this commensurability.

**3.4 The influence of the considered perturbations on the calculation results**

The influence of all eight large planets and, moreover, the relativistic effects of orbital perihelion displacement (REOPD) are not always taken into account in calculating evolution of small body orbits. It seems quite clear that where Mercury is taken into account, the REOPD must be included as well. But this holds true not only for Mercury. For the asteroids in the orbits with large eccentricities, i.e. with small perihelion distances, these effects should be also considered. Numerical calculations of the asteroid orbit evolutions were performed both with taking the REOPD into account and without. It was originally done for asteroid №506437. The initial eccentricity of the orbit is 0.72 ($q = 0.88$ AU). In the previous calculations, the asteroid passed to Centaurs after 14,000 years of the evolution. In calculations without the REOPD, the asteroid passed to Centaurs after 25,000 years. Therefore, the REOPD should be taken into account for such orbits.

For comparison, similar calculations were performed for an orbit outside the resonant zone and with moderate eccentricity and inclination ($a = 3.10$ AU, $e = 0.15$, $i = 10^o$). At different stages of the evolution, the differences in the orbit semimajor axis (with and without the REOPD) were: 20,000 yr: $2\times10^{-6}$ AU, 40,000 yr: $4\times10^{-6}$ AU, 60,000 yr: $1\times10^{-5}$ AU, 80,000 yr: $4\times10^{-5}$ AU, 100,000 yr: $2\times10^{-4}$ AU. The necessity of using the REOPD for such orbits depends on the problem to be considered.

In our calculations, where bodies could approach the planets, even very little differences in the orbit elements yielded noticeably different results. Therefore, again, there is a need to take the REOPD into account, especially since the duration of calculations does not change.

The situation is similar to the perturbations from the planets of the terrestrial group. Interestingly, some asteroids in our sample were able to pass to Centaurs without noticeable perturbations from the planets. We recalculated the orbit evolution for six asteroids that had passed to Centaurs, but we took into account the perturbations only from the four giant planets (Jupiter - Neptune). Even so, three asteroids passed to Centaurs, although slightly later than in the previous calculations. The other three remained in resonance for 100,000 years.

The Table demonstrates some orbit characteristics of these 6 asteroids: the initial semimajor axes values, eccentricities, inclinations and *ds* values. The asteroids that passed to Centaurs under perturbation of only 4 planets are marked with asterisks.

The Table shows that bodies pass to the Centaurs region, where they are in orbits with higher eccentricities ($e > 0.49$), and at that they also have higher *ds* values ($ds > 140^o$) and slightly lower inclinations. Such parameters enable them to go out of resonance and approach Jupiter without the terrestrial planets influencing such bodies. At the same time, these planets can speed up the exit from resonance.

*Table* The initial orbit elements of asteroids that passed to Centaurs

| Number | *a*, AU | *e* | *i*$^o$ | *ds*$^o$ |
|---|---|---|---|---|
| 322713* | 3.391 | 0.584 | 5.46 | 144 |
| 406803* | 3.331 | 0.495 | 8.90 | 158 |
| 506437* | 3.141 | 0.717 | 10.25 | 164 |
| 451124 | 3.206 | 0.699 | 10.02 | 98 |
| 471179 | 3.392 | 0.369 | 15.91 | 153 |
| 475265 | 3.196 | 0.428 | 17.37 | 170 |

Taken separately, the high *e* or *ds* values do not create conditions for transitions to the Centaur region in the absence of terrestrial planet perturbations. Orbits 475265 ($ds = 170^o$) and 451124 ($e = 0.70$) can be seen as the prime examples thereof. An asteroid approaching the terrestrial planets provide for an exit from resonance for such orbits. The exits from resonance of orbits 471179 and 475265 are probably due to the asteroid approaching Mars, although not too close. These orbits have sufficiently large initial values of *ds,* and their exits out of the resonance can occur with relatively little additional perturbations.

For asteroid №451124, the situation is slightly different. Its initial value of *ds* is $98^o$. However, taking into account perturbations from 8 planets, the orbit goes out of the resonance in 23,000- year time. Figures 8 and 9 show the $a(t)$ and $\sigma(t)$ dependences for this orbit until the very moment of the exit from resonance. A correlation between the amplitudes of oscillations of *a* and σ can be clearly visible. For comparison, the σ(*t*) dependence for the same orbit, taking into account perturbations of only 4 planets, is presented in Fig. 10. Here the value of *ds* varies within 94° – 100°. This dependence is given within the time interval of 23,000 yr, while the asteroid course does not change in the time interval of 100,000 yr.

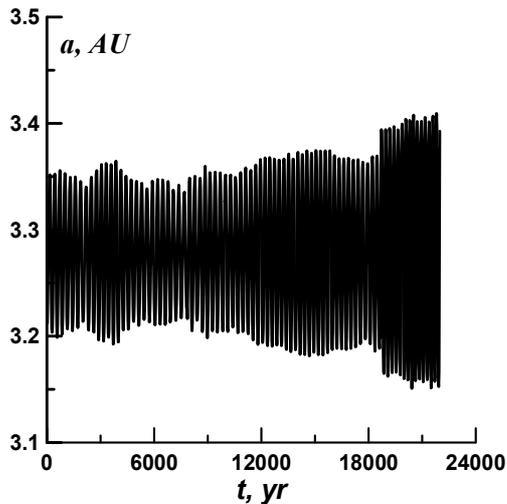

*Figure 8.* The *a*(*t*) dependence for the orbit 451124 (calculations with 8 planets)

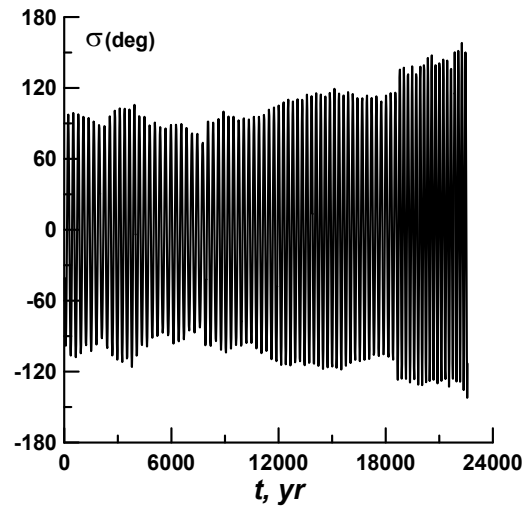

*Figure 9.* The σ(*t*) dependence for the orbit 451124 (calculations with 8 planets)

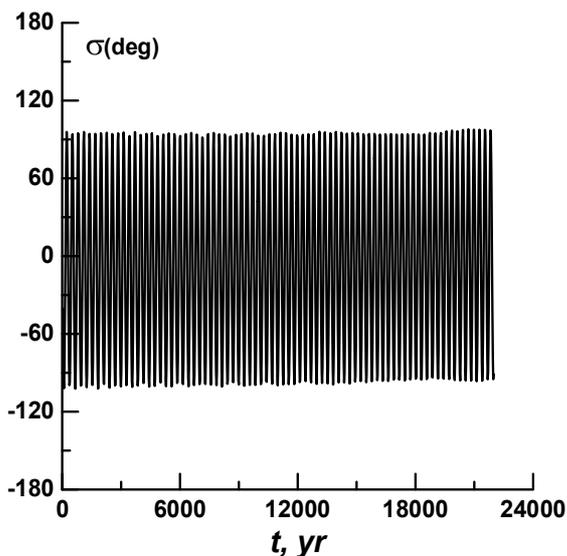

*Figure 10.* The σ(*t*) dependence for the orbit 451124 (calculations with 4 planets)

The difference between the σ(*t*) dependences in Fig. 9 and Fig. 10 can clearly indicate that the change of *ds* of the orbit 451124, as well as its exit from resonance, is very likely to have been influenced by the planets of the terrestrial group. The asteroid approaches to the Earth play the main part. For 22,000 yr, there have been more than 30 instances of the asteroid approaching the Earth at the distances which are smaller than 0.05 AU. Of these, two approaches were at the distances smaller than 0.01 AU. The first one was at the time *t* = 7,843 yr, and the other one – at the time *t* = 18,646 yr. As can be seen from Fig. 9, just after these approaches, the *ds* values began to change in a more significant manner. It is especially noticeable after the second approach: the *ds* value almost immediately increased from 112° to 132°. Then it gradually increased to 143°, and the orbit left the resonance.

Thus, in our case, we should take into account both the relativistic effects of the orbital perihelion displacement and the perturbations from all 8 planets.

## 4. DISCUSSION

In general, according to our estimates, 6 body orbits have passed from the 2:1 MMR with Jupiter and the bodies moved to the Centaur population over 100,000 years. Such a transition, however, does not cause any doubt only in two cases, which is definitely a small number. Probably, if we had calculated considerably longer time intervals (millions of years), the number of such transitions would be bigger. Yet, confirming the accuracy of such calculations will be problematic. The forward and backward method can be effective only at relatively short time intervals. The clone method can be used to estimate some averaged orbit characteristics of a certain type (life time, the dominant direction of evolution). But this method cannot accurately confirm the evolution of an individual orbit. In addition, we did not set out to detect all bodies in resonance that pass to Centaurs. The main task of this study is to show that in principle it is possible for bodies that are in the 2:1 commensurability with Jupiter to move to the Centaur region. Eventually, some of such bodies may be found in that area. And this consideration should be taken into account when the physical properties of the Centaur population are studied.

This study does not cover all features of orbit evolution in resonance. We do not analyse the mechanism of increasing eccentricities. The main thing is that such an increase does take place in eccentricities. We show the mechanism of how bodies exit from stable resonant zones. For some bodies, their exits from resonance became possible only after they had approached the terrestrial planets. Such approaches help orbits to go out of the resonance, even where *ds* values are rather small. Other bodies can go out of resonance and pass to Centaurs without the gravitational influence of the terrestrial planets. These are bodies that have large eccentricities, significant *ds* values and slightly smaller inclinations.

Nor does this paper not analyse the question of the life time of bodies in the 2:1 commensurability with Jupiter. Furthermore, this time significantly differs for bodies in different resonant orbits. Some asteroids may exist in the resonance for more than billions of years, some - for hundreds of millions, and others - for less than tens of millions years (Morbidelli 1996, Roig et al. 2002). According to our calculations, the bodies that passed to Centaurs obviously belonged to the latter group.

In our calculations, each irreversible exit of the orbit from resonance ended with the body approaching Jupiter and the transitioning to the Centaur region. We have not detected any cases of transition to the zone of objects with smaller semimajor axis values. However, the number of transitions to the Centaur region is low. Knowing this, we can conclude that if not the majority, then a significant part of bodies that had once left the 2:1 gap moved beyond the Jupiter orbit. Some of them may have left the Solar system long ago. And some may still be among Centaurs to this day.

## 5. CONCLUSIONS

Bodies from the 2:1 mean motion resonance with Jupiter can pass to the Centaurs population. Transitions to the Centaur region occur after reaching sufficient values of orbit eccentricities, where orbits depart from the resonance and bodies approach Jupiter. When bodies approach the terrestrial planets, it helps such bodies to escape resonance.

## DATA AVAILABILITY

The data underlying this article are available in the Minor Planet Center catalog and can be accessed with https://minorplanetcenter.net//